**An aptamer-biosensor for azole class antifungal drugs.**

Gregory R Wiedman, Yunan Zhao, Arkady Mustaev, Jinglei Ping, Ramya Vishnubhotla, A.T. Charlie Johnson, David S Perlin.

**This report describes the development of an aptamer for sensing azole antifungal drugs for therapeutic drug monitoring. Modified Synthetic Evolution of Ligands through Exponential Enrichment (SELEX) was used to discover a DNA aptamer recognizing azole class antifungal drugs. This aptamer undergoes a secondary structural change upon binding to its target molecule as shown through fluorescence anisotropy-based binding measurements. Experiments using circular dichroism spectroscopy, revealed a unique double G-quadruplex structure that was essential and specific for binding to the azole antifungal target. Aptamer-functionalized Graphene Field Effect Transistor (GFET) devices were created and used to measure the binding of strength of azole antifungals to this surface. In total this aptamer and the supporting sensing platform could provide a valuable tool for improving the treatment of patients with invasive fungal infections.**

Understanding a drug's pharmacokinetics is crucial to safely and effectively treating patients. Unfortunately, drug levels in patients can vary significantly, and the factors contributing to this variability are frequently misunderstood. For some critically-ill patients, it is essential to gauge levels of drug in real-time. The best therapeutic management can be achieved by maintaining a therapeutic level in a patient's bloodstream and by optimizing individual dosage regimens. These analyses generally rely upon trough and peak monitoring, and real-time kinetic drug modeling. For this reason, therapeutic drug monitoring (TDM) of some drugs is a critical component of successful therapy (1). It is particularly important to monitor drugs with narrow therapeutic ranges, marked pharmacokinetic variability, target concentrations that are difficult to monitor, and drugs known to cause adverse events.

The azole antifungal drugs posaconazole, fluconazole, voriconazole, and itraconazole, are an important class of lanosterol 14α-demethylase enzyme inhibiting molecules (2), which compromise fungal cell membranes by preventing the synthesis of the key component ergosterol (3). A number of these drugs are highly hydrophobic, which creates analytical challenges. Furthermore, because of their hydrophobic nature it is difficult to know how much of the drug is freely available in the blood at any given time. Wide variances in the pharmacokinetics of critically ill patients have been observed for triazole drugs like voriconazole and posaconazole, which has resulted in a need for TDM (4). Furthermore, posaconazole and voriconazole have been shown to have drastically different bioavailability depending on how they are administered and if they are co-administered with other drugs (5, 6). Therapeutic drug monitoring in conjunction with antifungal therapy has been shown to promote more favorable outcome compared to the non-TDM group (7). Unfortunately, TDM requires blood to be drawn from patients and then drug levels in blood evaluated by analytical instrumentation, at some later point in time. Analytical techniques using machinery such as Liquid Chromatography and Mass Spectrometry require skilled staff and resources that are not found in all hospitals (8). These barriers become especially difficult to overcome when treating patients in community hospitals, at home or in resource limited settings. Effective methods for sensing small drug molecules in blood samples would make it easier to determine drug concentrations.

Any effective TDM method requires a way to capture the drug target from a patient sample. Antibodies provide specificity and sensitivity as a capture probe, but they are typically unstable over a wide range of assay conditions. As a more robust alternative, oligonucleotide-based aptamer capture probes were developed as a stable and selective capture molecule for small molecule drugs. Oligonucloetide aptamers can bind to a wide variety of target molecules with high affinity. Such oligonucleotides, i.e. DNA and RNA libraries of $10^{14}$-$10^{16}$ molecules can be quickly synthesized and screened using the *in vitro* SELEX (Synthetic Evolution of Ligands through Exponential Enrichment) methods (9, 10).

In this report, azole-specific aptamer were created, which bind posaconazole in its free form by using a modified SELEX method to screen a library of more than $10^{14}$ DNA sequences. Furthermore, Graphene Field Effect Transistors (GFETs) were developed as a biosensing platform for detecting azole antifungal drugs with these aptamers. These devices represent a newly emerging type of biosensor that relies on changes in current and gate voltage of the transistor itself rather than the flow of an electrolyte solution, binding of antibodies, or fluorescence labeling (11). Taken together these results provide a possible path forward for development of an azole antifungal sensing device with potential broader downstream capability of improving therapeutic drug monitoring of small molecule drugs.

**Results**

**SELEX Process Results.** A posaconazole-binding aptamer was generated from a random 40-mer library using a modified SELEX process. PCR output and Oligreen Dye intensity were used to track the enrichment of posaconazole binders. The output of the PCR experiments for each selection round was plotted after 15 and 25 cycles of PCR amplification in addition to the intensity of Oligreen dye on DNA containing counter SELEX and SELEX (target) beads ( Fig. 1 A,B). During rounds of increased pressure such as round 5 and 6, the total PCR output decreases, but then recovers in later rounds. This is indicative of enrichment of target binding aptamers. Furthermore, the relative amount of aptamer bound to target beads versus control beads increases in later rounds. This becomes especially clear past Round 10 when the beads are first washed with the competitive molecule fluconazole. This wash allows for weak binding molecules to be eluted from the sample.

Denaturing polyacrylamide gels were used to assess whether or not their molecular weight changed during SELEX. Each round displays two bands, which are a result of leftover double stranded DNA not digested by λ exonuclease ( Fig. 2A). The single strand bands from Rounds 1-10 are the same size as the single strand control library. The larger bands from Rounds 11, 12, and 13 are still distinct from the library band when denatured ( Fig. 2C). Each band was separated as described in the Methods section and further analyzed.

**Selection of Best Consensus Sequence.** After separation, the "Heavy" and "Light" bands were analyzed separately for their binding capabilities. The Light bands of Round 12 and 13, those the same length as the library, do not bind to the target (Fig. 3). The Heavy bands, however, bound significantly better to the posaconazole labeled beads versus the control beads and were used for further studies. The heavy

| Rd 13 | CGGGGGGAGGCGGAGGGAGGACTGGGGCTTCATTGACGTTCTTCACAGTAGGGGTAAGGGCTTAGGTGGTTGGTGCCTG |
| Rd 9 | CGCGGGGAGGAGGAGGCAGACTGGGGCTTCTTTGACGTTC |

bands from Round 13 as well as the sample from Round 9 were sequenced.

Table 2

Here, red indicates overlap and yellow highlight shows regions of predicted G-quadruplex structure. The consensus sequence for Rd 9 is the same size, 40 bases, as the original library. The consensus sequence for Rd 13 is almost double that size at 79 bases. A significant portion of the Rd 9 sequence, in red, is found in the Rd 13 sequence. Additionally, QGRS mapping was used to predict the existence of G-quadruplex structure in these sequences (12). The Rd 9 consensus sequence contains one predicted G-quadruplex stretch, highlighted. Interestingly, the Rd 13 sequence contains two regions of predicted G-quadruplex structure. Stem-loop structure predicting software was used to map the room temperature (298 K) structures of these aptamers (S2) (13). These predictions show that while Rd 9 forms a stem-loop structure with a single arm, Rd 13 has two separate arms. These features became a major focus of the further study of these aptamers.

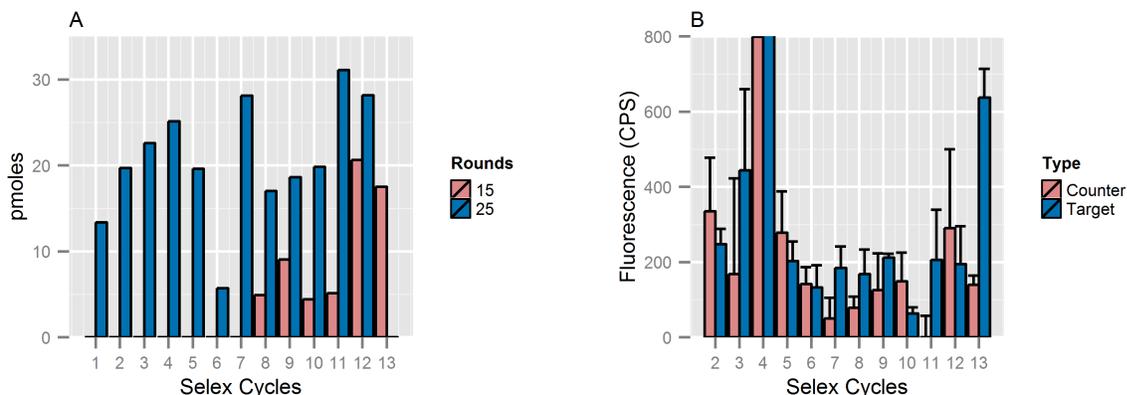

Fig. 1. (A) the number of pmoles of PCR output after each round. After rounds 5 and 6 the output decreases due to increased pressure. The output increases in later rounds and amplification can even be seen after only 15 cycles of PCR as opposed to 25 cycles. In (B) the fluorescence intensity of the Oligreen dye from aptamer samples incubated with either control beads or posaconazole (target) labeled beads. In later rounds the amount bound to the target beads generally increases as binders are enriched

**Control Sequences.** This study included several control sequences to further investigate the importance of the structure of the Rd 13 aptamer: Rd 13 Scrambled (S), Rd 13 T6 and Rd 13 T1 were created with varying middle lengths.

| Rd 13 S | GCTGTGTGCGTGAGTGGAGTGCGCGCGAGAGTGAGT GCGGTGAGAGATTCGGTGTGCGTGTCGTGATGAATGC GACCGG |
|---|---|
| Rd 13 T6 | CGGGGGGAGGCGGAGGGAGGACTGGGGCTGTAGGG GTAAGGGCTTAGGTGGTTGGTGCCTG |
| Rd 13 T1 | CGGGGGGAGGCGGAGGGAGGACTGGGTGGGGTAAG GGCTTAGGTGGTTGGTGCCTG |

Table 3

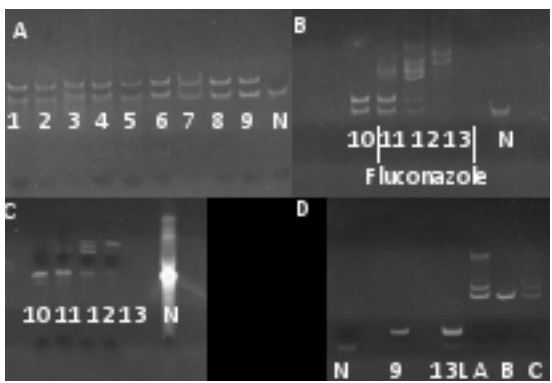

Fig. 2 After the addition of fluconazole there are bands that are higher in molecular weight than the starting library N and the original sequences. A denaturing polyacrylamide gel (C) highlights the fact that these bands are, in fact, unique molecular weight aptamers and not aggregates of the smaller original library sequence N.

Rd 13 Scrambled, was carefully reordered from the original sequence to fully prevent the formation of G-quadruplex structures. The two other sequences maintained the G-quadruplex regions but with two different distances between them.

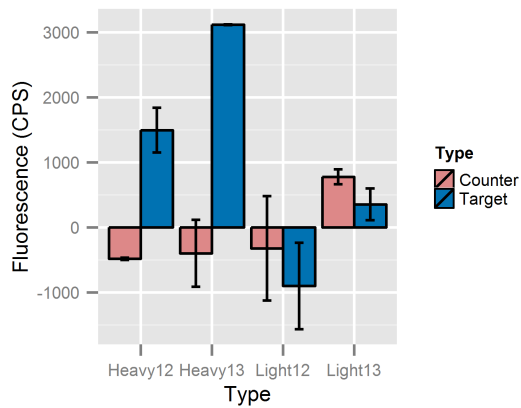

Fig. 3 The relative binding capabilities of the heavy and light bands are different to Control (Counter) and Target (Posaconazole) labeled beads respectively. Light bands from Round 12 and Round 13 showed little preference for either type of beads. The Heavy bands, however, bound significantly better to the target labeled beads as compared to the control beads.

**Binding Affinity of Azole Drug Aptamers.** Small molecule aptamers work best when they have strong binding affinity and low dissociation constants. Fluorescence anisotropy experiments were used to determine the dissociation constants for the posaconazole-aptamer complex (14, 15). In these experiments only the Round 13 aptamer caused an increase in anisotropy from titrations of a constant PosBD concentration with increasing amounts of DNA aptamers (Fig. 5A). The control library and Rd 9 fail to bind to the PosBD. It should be noted that the control library contains the primer sequences, bringing the total length to 86 nucleotides. The difference in molecular weights between Rd13 and Rd9 is not the sole reason for their difference in anisotropy. The same data can be generated by holding the DNA aptamer concentration constant and titrating with increasing amounts of PosBD (Fig. 5B). The dissociation constants for each experiment are 2.7 ± 1.2 µM and 0.98 ± 0.18 µM, respectively. The overall dissociation constant for PosBD might be weaker than that of posaconazole given the fact that the latter was used in the SELEX process and not PosBD. The differences in anisotropy changes were further used to probe the specificity of the aptamer for the target. Specificity was interrogated with respect to two other BODIPY labeled molecules: Isavuconazole (ISV), which is chemically similar to posaconazole, and caspofungin (CSF), which is chemically dissimilar since it is an echinocandin class drug. Titrating 100 pmoles of Rd 13 aptamer into a 125 µL solution containing 100 pmoles of PosBD causes a greater anisotropy change than titrating into 100 pmoles of ISV_BD or CF_BD (Fig. 6). Titrating Rd9 or the Library into 100 pmoles of PosBD causes little to no change in anisotropy. The truncated versions T6 and T1 cause less of an anisotropy change, though Rd13 T6 is not significantly different (Fig. 6 B). The Rd 13 scrambled aptamer did not cause a significant anisotropy change compared to a control such as EDTA and hence did not bind to PosBD. These results indicated that the G-quadruplex structure is necessary for binding.

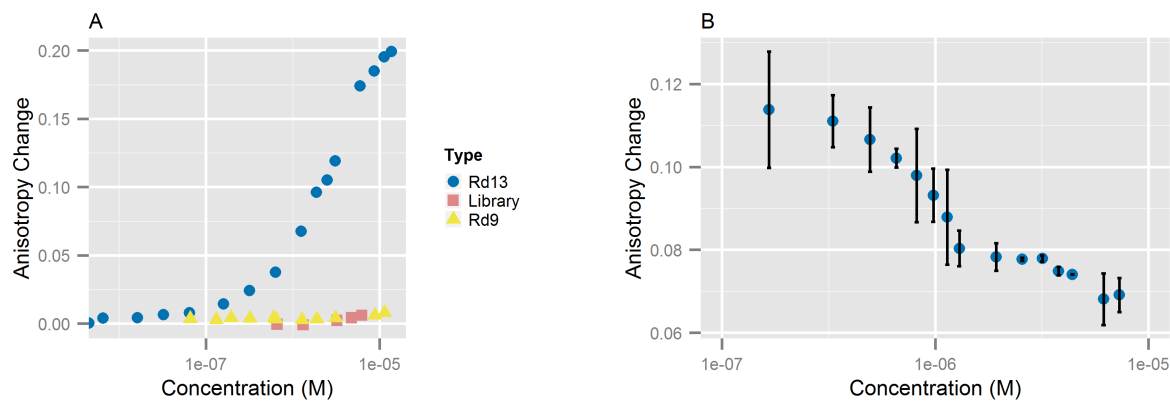

**Fig. 5** The anisotropy changes when aptamer was titrated into a 100 pmoles/125 µL of PosBD. The anisotropy changes neither with the earlier round (Rd 9) nor with the control library. (B) data from titrations of 1000 pmoles Rd13 in 125 µL modified SELEX buffer with increasing amounts of PosBD. The dissociation constants from each experiment were calculated to be 2.7 ± 1.2 µM and 0.98 ± 0.18 µM respectively.

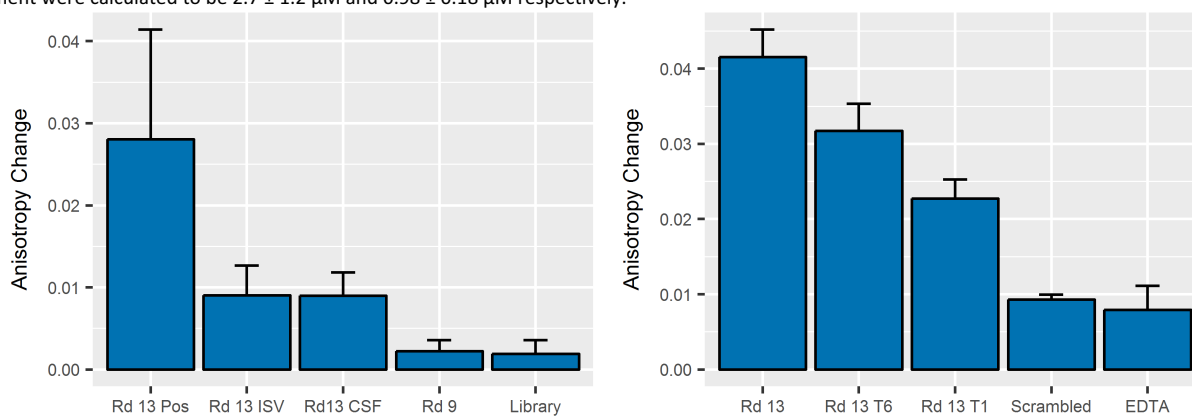

**Fig. 6** 100 pmoles of aptamer (Rd 13, Rd 9, or the Library) were titrated into 100 pmoles of varioius BODIPY labeled drugs (Posaconazole (Pos), Isavuconazole ISV), and Caspofungin (CSF) (A). Titrating Rd 13 into PosBD caused a significant change in anisotropy. Titrating Rd 13 into other BODIPY labeled drugs or Rd 9/Library into PosBD caused little changes in anisotropy. These data suggest that Rd 13 binds best to Posaconazole. Different length aptamers were titrated into 100 pmoles of PosBD (B). The full length Rd 13 aptamer caused the greatest change in anisotropy and the truncated and scrambled versions showed decreased amounts of anisotropy change. The scrambled version only caused a change proportional to that caused by a chelator, EDTA.

**Binding Competition Assay.** The aptamers developed in this study bind to azole class antifungal drug targets, specifically to those with the exposed terminal azole group like posaconazole. High specificity is important for downstream diagnostic devices to prevent false positive readings. The anisotropy experiments were modified slightly to develop a competitive assay to further probe specificity. This experiment showed the relative ability of various drugs to replace PosBD in the aptamer complex (Fig. 7). As expected, Posaconazole displaces the greatest amount of PosBD. Related drugs Fluconazole and Itraconazole replace fewer PosBD molecules. The chemically distinct Echinocandin antifungal drugs Micafungin and Caspofungin had little effect. Of the azole drugs, Itraconazole is the most hydrophobic. The lower amount of PosBD replacement with Itraconazole versus Posaconazole suggest that binding is not solely driven by hydrophobic effects. Secondary structure plays a large part in the binding of azole targets to these aptamers.

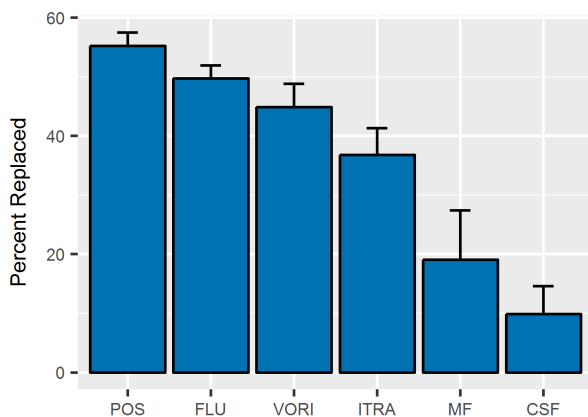

Fig. 7 A 50 pmole/125 µL solution of PosBD was incubated with 1000 pmoles of Rd 13 aptamer. After equilibration the aptamer was heated at 70C in the presence of Posaconazole (Pos), Fluconazole (Flu), Itraconazole (Itra), Voriconizole (Vori), Micofungin (MF), or Caspofungin (CSF) and then cooled on ice. The percent replaced was calculated as a function of the loss in anisotropy.

**Circular Dichroism and Secondary Structure Analysis.** Secondary structure controls the complexing of aptamers and target molecules. As these aptamers all contained multiple stretches of guanine residues, there is a high possibility that they form G-quadruplex structures. CD Spectroscopy experiments were conducted to probe the folding of these aptamers in the presence of salts and posaconazole. The CD spectra of Rd13 are characteristic of G-quadruplex folded DNA with a maximum at 260 nm and a minimum at 240 nm (Fig. 8 A) (16, 17). Addition of magnesium chloride to the solution both increases the signal at 260 nm and decreases the signal at 240 nm. This suggests that the aptamer forms a G-quadruplex structure in low salt buffer, which is slightly enhanced with the addition of salts. In contrast, the signal is not altered significantly by adding posaconazole in the absence of divalent salts (Fig. 8 B). The most drastic change occurs when the aptamers are exposed to a combination of both posaconazole and salts. The aptamers Rd 9 and Rd 13 show similar CD signals in 0.2 mM magnesium chloride. A G-quadruplex structure formed with Rd 9 and Rd 13 but not with Rd 13 scrambled (Fig. 8 C). When 100 µM posaconazole was added there is change in the CD signal for Rd 9 and Rd 13. With posaconazole, the spectra for Rd 9 changes to contain a maximum at 230 nm and a drastic minimum at 280 nm (Fig. 8 D). The spectra for Rd 13 in magnesium chloride with posaconazole contains two maximum at about 230 nm and 270 nm with a minimum above 300 nm. The Rd 13 scrambled sequence does not undergo any further change in secondary structure. The G-quadruplex structure of Rd 13 forms in the presence of salt and that this structure then changes when the target is added.

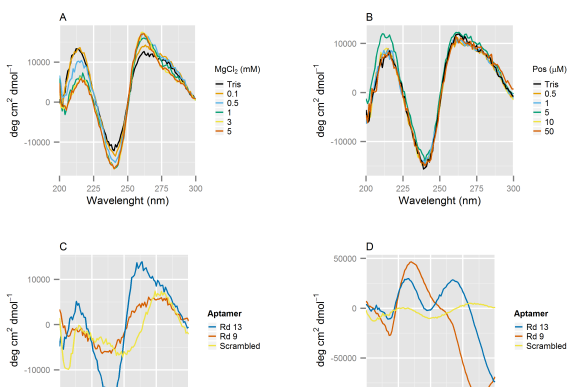

Fig. 8 Addition of magnesium chloride to the Rd 13 aptamer enhances G-qudruplex folding (A) but addition of posazonazole does not (B). This is seen in an increase in the maximum at 260 nm and a decrease in the minimum at 240 nm. Both Rd 13 and Rd 9 exhibit G-quadruplex structure in 0.2 mM Magnesium chloride salt (C). The structure changes drastically in the presence of magnesium chloride and 100 µM posaconazole (D). The Rd 13 scrambled aptamer does not change but Rd 9 now contains a peak 230 nm and a minimum at 280 nm and Rd 13 contains two peaks at 230 nm and 260 nm with a new minimum peak above 300 nm

**Graphene Field Effect Transistors.** When incorporated with Graphene Field Effect transistors, these aptamers act in an induced-fit manner, which allows them to function as a small molecule capture arm on a supported surface. GFET devices functionalized with amino-Rd13 were used to measure Posaconazole concentration (Fig. 9). Posaconazole was diluted from DMSO into SELEX buffer as described above. As concentrations of Posaconazole were increased from 0.01 µg/mL to 100 µg/mL, the gate threshold voltage shift increased to upwards of -6V. In contrast, treatment with the echinocandin drug Caspofungin produced a negligible shift in the gate threshold voltage. These values are in the range of was has been seeing with antibody-functionalized biosensing devices and a related biosensor for HIV drugs(18, 19). Interestingly, the aptamers on these devices were shown to have a dissociation constant of 1.8 ± 0.5 µg/mL or approximately 1.8 µM. These values are in good agreement with the dissociation constant derived from the anisotropy assay. These data suggest that support functionalized Rd13 aptamer binds posaconazole in a similar fashion to free Rd13 aptamer.

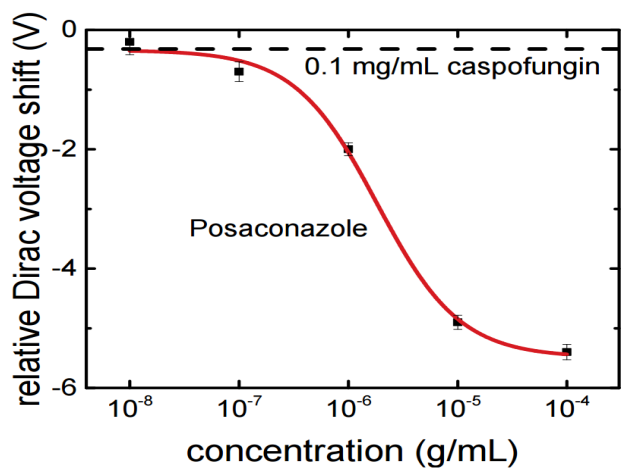

Fig. 9 Data of the response of Rd13 functionalized GFET devices to azole class drug Posaconazole and echinocandin class drug Caspofungin. The devices showed a detectable threshold, Dirac Voltage shift between 0.1 mg/mL and 0.1 µg/mL. These values, especially the low µg/mL range are therapeutically relevant concentrations of azole class antifungal drugs. Notably, even at the highest concentration, 0.1 mg/mL caspofungin caused no detectable response above the background, indicating specificity for azole antifungal drugs.

**Discussion**
Therapeutic drug monitoring requires a method of capturing molecules and separating them from a sample for analysis. This report highlights the development of a drug capturing oligonucleotide using our SELEX process and the discovery of a unique structure, which allows it to bind azole antifungal drugs. Circular Dichroism spectrophotometry shows that this oligonucleotide works as a scaffold with two sections of G-quadruplex folds. Large protein target aptamers have been made before using two, separate G-quadruplexes linked chemically to bind at separate sites (20). These types of folds rarely interact with smaller hydrophobic molecules due to the highly charged nature of single strand DNA. The Rd13 aptamer was generated as a result of single G-quadruplex aptamers of this library interacting directly. These are not simple stem-loop structures found in most SELEX assays. Instead this library and SELEX process managed to access potential quaternary structures by overlapping G-quadruplexes. When the target binds to these aptamers the quaternary structure can transform from overlapping G-quadruplex to another structure similar to a B->Z DNA transition(21, 22). This transition can play a role in developing downstream applications.

This type of dual G-quadruplex aptamer is ideal for capturing small hydrophobic molecules. The anisotropy binding experiments show that the G-quadruplex is essential for target binding. When bound to the surface of a Graphene Field Effect Transistor, the aptamer works as a capture arm. This arm collects posaconazole from the sample, which leads to a change in the GFET gate voltage. The azole "aptasensor" adds to the list of other aptamer-based sensing devices (23, 24). Unlike these other devices, however, GFET aptamer biosensors have the potential to function without the need for secondary antibodies, fluorophores, or electrochemical mediators. The versatility of oligonucleotide based biosensor opens the door to numerous different applications.

Taken together, the unique structure and binding properties of Oligonucleotide provided with the sensitivity of Graphene Field Effect Transistors could prove useful for therapeutic drug monitoring. Posaconazole and the other azole class antifungal drugs exhibit strong hydrophobicity and protein binding. Despite this fact, this aptamer binds specifically to azole drugs. There are other aptamers towards clinically interesting drugs as well such as the Aminoglycoside Antibiotics and Antiretroviral Drugs among others(25, 26). The work presented here provides the first basic step towards effective Therapeutic Drug Monitoring: a method of capturing and sensing the drug. These other aptamers could also be utilized as capture arms for a graphene-based sensing platform. The next step will involve testing patient samples and validating their usefulness in clinic. Such a devices will allow clinicians to quickly assess azole concentrations in a patient's blood and to provide them with the additional care that they need.

**Materials and Methods**
N40 DNA Aptamer library was purchased from Tri-link Biotechnologies Inc (Ronkonkoma, NY). Other oligonucleotides including amino-functionalized oligonucleotides were synthesized by IDT (Coralville, IA), Sigma-Aldrich (St. Louis, MO) and Biosearch Technologies (Novato, CA). Carboxyl Dynabeads (14305D), Oligreen ssDNA assay kit, and BODIPY FL Dye were purchased from Life Technologies Co (Carlsbad, CA). Streptavidin magnetic beads and Lamda Exonuclease were purchased from New England Biolabs (Boston, MA). Azole Drugs were purchased from Santa Cruz Biotechnologies (Dallas, TX). All other reagents and solvents were purchased from Thermo Fisher Scientific (Waltham, MA). Graphene devices were fabricated in-house using methods described in previous work(11)

**SELEX Process.** The aptamer library from Trilink Biotechnologies was prepared by dissolving 1 nmol of DNA in 100 µL of SELEX buffer (140 mM Sodium Chloride, 2 mM Potassium Chloride, 5 mM Magnesium Chloride, 2 mM Calcium Chloride, 0.05% Tween in 20 mM pH 7.4 Tris Buffer). The library was heated at 94°C for 3 minutes, placed on ice for 5 minutes, and then incubated at room temperature (25°C) for 5 minutes. Next, the DNA library was incubated for 1 hour at 50°C and 2 hours at room temperature, then in later rounds 10 minutes at 50°C and 20 minutes at room temperature. For the first round, the library was incubated with 1 mg of unlabeled carboxyl dynabeads (counter SELEX). The beads were carefully concentrated using a magnet. The library was then incubated with 1 mg of posaconazole labeled beads for 1 hour at 50°C and 2 hours at room temperature. The beads were washed 3 times with 100 µL SELEX buffer and 1 time with 100 µL Millipore water incubated with 1 nmol posaconazole in 20uL of water with 0.01% DMSO for 1 hour. The recovered DNA was purified using a Zymo Research DNA preparation columns. Recovered DNA was amplified in two Polymerase Chain Reaction (PCR) conditions. First, the 20 µL of DNA was amplified using a Takara rTaq DNA polymerase enzyme. The mixture contained, 2.5 µL of (10 µM) Forward Primer, 2.5 µL of (10 µM) Phosphorylated Reverse Primer, 5 µL of 2.5mM dNTPs, 5µL of 10X MgCl2 buffer, 0.5 µL (2.5 units) of rTaq Enzyme, 20 µL DNA and 14.5 µL Millipore water. PCR was performed with the following conditions: 94°C 5 minutes initial then 15 cycles of 94°C for 30 seconds, 50°C for 30 seconds and 72°C for 1 minute followed by a final 5 minutes of extension at 72°C. This sample was then treated as-is with 1 µL (5 units) of λ Exonuclease and incubated at 37°C for 30 minutes. This mixture was purified using a Zymo columns and the output was checked for absorbance at 260 nanometers using a Nanodrop spectrophotometer. If the yield was less than 1pmol an additional PCR was conducted as above but substituting 10 cycles for the 15 cycles. Further rounds of SELEX included the following modifications: For round 2 through 5, 200 pmoles capacity beads were used instead of 1 nmoles. After round 5, incubation times were decreased to 30 min counter SELEX, then 10 minutes at 50°C and 20 minutes at room temperature. After round 6 the bead capacity was decreased to 20 pmoles. In rounds 11 to 13 beads were incubated for 1 hour with 1 µL of 1M Fluconazole in 20 µL first as an additional competitive wash before washing with Posaconazole. (S1)

**SELEX Results Tracking.** In each round 1 µL of DNA-incubated control beads and 1 µL of DNA-incubated posaconazole labeled beads were saved after washing with water but before posaconazole elution. The DNA content was assessed using a 1:800 solution of Oligreen dye in 20 mM Tris buffer with 2 mM EDTA at pH 7.5 The sample were analyzed using a Photon Technology International (PTI) Fluorometer. PCR output was measured using a Nanodrop Spectrophotometer. DNA size was investigated by running an 8% polyacrylamide gel and denaturing gels were run using 8% polyacrylamide gel with 8 M Urea in SDS buffer after loading DNA treated with formamide at 100 °C. Sequencing of various rounds was performed by Macrogen (Rockville, MD) to determine the consensus sequence for that round.

**Fluorescence Binding Experiments.** Purified bands and synthesized consensus sequences were prepared by taking 2 pmoles and dissolving them in 100 µL SELEX buffer. These solutions were heated at 94°C for 3 minutes, on ice for 5 minutes, and at room temperature for 5 minutes. The samples were incubated with 20 pmol capacity posaconazole labeled and unlabeled beads at 50°C for 10 minutes and room temperature for 20 minutes. The samples were then washed 2 times with water and 125 µL of a 1:800 dilution Oligreen Dye was added. Samples were then heated at 94°C for 3 minutes, the beads were concentrated, and the supernatant was collected. Samples were excited at 480 nm with emission scanning from 500 nm to 550 nm. The fluorescence was recorded as CPS at 520 minus the CPS at 505 nm.

**Fluorescence Anisotropy Binding Experiments.** Fluorescence Anisotropy experiments were conducted using a PTI Fluorometer with Fluorescence Polarizers. One hundred pmoles of Posaconazole BODIPY (PosBD) was added from DMSO (1 µL) to 125 µL of modified SELEX buffer (140 mM Sodium Chloride, 2 mM Potassium Chloride, 5 mM Magnesium Chloride, 2 mM Calcium chloride, in 20 mM pH 7.4 Tris Buffer)(12). Fluorescence anisotropy experiments were recorded using a polarizer system and the G-factor was calculated manually for each run but consistently fell within 0.44-0.45. Anisotropy measurements were recorded first for 2 minutes. After the initial 2 minutes, aliquots of aptamers from 1 pmoles to 2000 pmoles were added, and samples were equilibrated for 5 minutes. The value of the anisotropy was taken to be the average anisotropy of the last 60 seconds after equilibration. Anisotropy values were plotted as the change in anisotropy:

**Equation 1:** $\Delta\langle r\rangle(C) = \langle r\rangle(C) - \langle r\rangle_o$

These values were used to calculate a bound fraction:

**Equation 2:** $F_{bound}(C) = \frac{\Delta\langle r\rangle(C)}{\Delta\langle r\rangle_{max}}$

The bound fraction was further used to calculate a dissociation constant by fitting to:

**Equation 3:** $F_{bound}(C) = \frac{C}{C+K_d}$

**Fluorescence Anisotropy Competition Assays.** Competition assays were performed using the same measurement techniques as described above for binding assays. In this experiment, 50 pmoles of PosBD was added from DMSO into 125 µL of modified SELEX buffer, and the anisotropy was recorded for 2 minutes. One thousand pmoles of aptamer was then added and allowed to equilibrate for 10 minutes. After this time, 1000 pmoles of an unlabeled drug molecule was added and the solution was heated up to 70oC for 3 minutes and cooled on ice for 2 minutes. The heat-ice cycle was performed twice. The anisotropy was then recorded again for another 5 minutes. The percent of PosBD replaced was calculated as:

**Equation 4:** $\% \, Replaced = 100 * \frac{\langle r\rangle_{aptamer,drug} - \langle r\rangle_0}{\langle r\rangle_{aptamer} - \langle r\rangle_0}$

Percent replaced equals 100 multiplied by the anisotropy with aptamer and drug replacement minus the initial anisotropy divided by the anisotropy caused by the aptamer alone minus the initial anisotropy

**Circular Dichroism Experiments.** Experiments were performed using an Aviv Model 420 CD Spectrophotometer. All aptamer samples were prepared at 10 µM concentration in 20 mM Tris buffer pH 7.4. Increasing amounts of salts and/or azole antifungal drugs were added and the CD spectra was recorded from 300 nm to 200 nm.

**Graphene Field Effect Transistors Functionalization and Testing.** Chemical Vapor Deposition (CVD) method with a methane source was used to grow graphene, which was then transferred via electrolysis onto a patterned Si/SiO2 surface. This surface contained chromium and gold electrodes and the graphene channels between them were further defined through photolithography and annealed in an argon/hydrogen environment. In order to functionalize the surface, GFETs were incubated in 1-Pyrenebutyric acid N-hydroxysuccinimide ester (P-base) and DMF for 20 hours. After this incubation, the devices were further incubated in a solution of PBS (pH 7.6) containing the Rd13 aptamer for 3 hours. The devices were heated from 70oC to 90oC, held at this temperature for 15 minutes, and then allowed to cool down to room temperature. In order to test the devices, the I-Vg properties of GFETs were measured dry pre- and post-incubation with either the azole drug posaconazole or the echinocandin drug caspofungin.

1. Andes D, Pascual A, & Marchetti O (2009) Antifungal Therapeutic Drug Monitoring: Established and Emerging Indications. *Antimicrob Agents Ch* 53(1):24-34.
2. Sheehan DJ, Hitchcock CA, & Sibley CM (1999) Current and emerging azole antifungal agents. *Clinical microbiology reviews* 12(1):40-79.
3. Ghannoum MA & Rice LB (1999) Antifungal agents: mode of action, mechanisms of resistance, and correlation of these mechanisms with bacterial resistance. *Clinical microbiology reviews* 12(4):501-517.
4. Jager NG, van Hest RM, Lipman J, Taccone FS, & Roberts JA (2016) Therapeutic drug monitoring of anti-infective agents in critically ill patients. *Expert Rev Clin Pharmacol*:1-19.
5. Dolton MJ, Bruggemann RJ, Burger DM, & McLachlan AJ (2014) Understanding variability in posaconazole exposure using an integrated population pharmacokinetic analysis. *Antimicrob Agents Chemother* 58(11):6879-6885.
6. Dolton MJ, Mikus G, Weiss J, Ray JE, & McLachlan AJ (2014) Understanding variability with voriconazole using a population pharmacokinetic approach: implications for optimal dosing. *The Journal of antimicrobial chemotherapy* 69(6):1633-1641.
7. Park WB, *et al.* (2012) The effect of therapeutic drug monitoring on safety and efficacy of voriconazole in invasive fungal infections: a randomized controlled trial. *Clin Infect Dis* 55(8):1080-1087.
8. Grebe SK & Singh RJ (2011) LC-MS/MS in the Clinical Laboratory - Where to From Here? *The Clinical biochemist. Reviews / Australian Association of Clinical Biochemists* 32(1):5-31.
9. Ellington AD & Szostak JW (1990) In vitro selection of RNA molecules that bind specific ligands. *Nature* 346(6287):818-822.
10. Tuerk C & Gold L (1990) Systematic evolution of ligands by exponential enrichment: RNA ligands to bacteriophage T4 DNA polymerase. *Science* 249(4968):505-510.
11. Ping J, Vishnubhotla R, Vrudhula A, & Johnson AT (2016) Scalable Production of High-Sensitivity, Label-Free DNA Biosensors Based on Back-Gated Graphene Field Effect Transistors. *ACS nano* 10(9):8700-8704.
12. Tucker WO, Shum KT, & Tanner JA (2012) G-quadruplex DNA aptamers and their ligands: structure, function and application. *Current pharmaceutical design* 18(14):2014-2026.
13. Reuter JS & Mathews DH (2010) RNAstructure: software for RNA secondary structure prediction and analysis. *BMC bioinformatics* 11:129.
14. Lea WA & Simeonov A (2011) Fluorescence polarization assays in small molecule screening. *Expert opinion on drug discovery* 6(1):17-32.
15. Jing M & Bowser MT (2011) Methods for measuring aptamer-protein equilibria: a review. *Analytica chimica acta* 686(1-2):9-18.
16. Paramasivan S, Rujan I, & Bolton PH (2007) Circular dichroism of quadruplex DNAs: applications to structure, cation effects and ligand binding. *Methods* 43(4):324-331.

17. Nagatoishi S, Tanaka Y, & Tsumoto K (2007) Circular dichroism spectra demonstrate formation of the thrombin-binding DNA aptamer G-quadruplex under stabilizing-cation-deficient conditions. *Biochemical and biophysical research communications* 352(3):812-817.
18. Lerner MB, Dailey J, Goldsmith BR, Brisson D, & Johnson ATC (2013) Detecting Lyme disease using antibody-functionalized single-walled carbon nanotube transistors. *Biosens Bioelectron* 45:163-167.
19. Vishnubhotla R*, et al.* (Scalable Biosensors Based on CVD Graphene for the Detection of the HIV Drug Tenofovir. *Submitted*.
20. Hasegawa H, Sode K, & Ikebukuro K (2008) Selection of DNA aptamers against VEGF165 using a protein competitor and the aptamer blotting method. *Biotechnology letters* 30(5):829-834.
21. Pohl FM & Jovin TM (1972) Salt-induced co-operative conformational change of a synthetic DNA: equilibrium and kinetic studies with poly (dG-dC). *Journal of molecular biology* 67(3):375-396.
22. Baker ES & Bowers MT (2007) B-DNA helix stability in a solvent-free environment. *Journal of the American Society for Mass Spectrometry* 18(7):1188-1195.
23. Hu R*, et al.* (2015) DLISA: A DNAzyme-Based ELISA for Protein Enzyme-Free Immunoassay of Multiple Analytes. *Anal Chem* 87(15):7746-7753.
24. Liu JW, Cao ZH, & Lu Y (2009) Functional Nucleic Acid Sensors. *Chem Rev* 109(5):1948-1998.
25. Stoltenburg R, Nikolaus N, & Strehlitz B (2012) Capture-SELEX: Selection of DNA Aptamers for Aminoglycoside Antibiotics. *Journal of analytical methods in chemistry* 2012:415697.
26. Kammer MN*, et al.* (2014) Characterizing aptamer small molecule interactions with backscattering interferometry. *The Analyst* 139(22):5879-5884.